# Energy Efficient Task Assignment in Virtualized Wireless Sensor Networks

Vahid Maleki Raee
*Concordia University*
Montreal, Canada
v_maleki@encs.concordia.ca

Diala Naboulsi
*Concordia University*
Montreal, Canada
d_naboul@encs.concordia.ca

Roch Glitho
*Concordia University*
Montreal, Canada
glitho@encs.concordia.ca

*Abstract*— Wireless Sensor Networks (WSNs) are being used extensively today in various domains. However, they are traditionally deployed with applications embedded in them which precludes their re-use for new applications. Nowadays, virtualization enables several applications on a same WSN by abstracting the physical resources (i.e. sensing capabilities) into logical ones. However, this comes at a cost, including an energy cost. It is therefore critical to ensure the efficient allocation of these resources. In this paper, we study the problem of assigning application sensing tasks to sensor devices, in virtualized WSNs. Our goal is to minimize the overall energy consumption resulting from the assignment. We focus on the static version of the problem and formulate it using Integer Linear Programming (ILP), while accounting for sensor nodes' available energy and virtualization overhead. We solve the problem over different scenarios and compare the obtained solution to the case of a traditional WSN, i.e. one with no support for virtualization. Our results show that significant energy can be saved when tasks are appropriately assigned in a WSN that supports virtualization.

## I. INTRODUCTION

Wireless Sensor Networks (WSNs) [1] are the key elements of the Internet of Things (IoT). They are being used more and more in multiple application domains. Traditionally, applications are embedded in WSNs; preventing a WSN to be used by several applications and usually leading to the deployment of redundant WSNs [2]. WSN Virtualization can help to address this issue. It enables the abstraction of physical sensing capabilities into logical sensing capabilities [2].

WSN virtualization can be done at node and at network levels. Node level virtualization allows multiple applications to run concurrent tasks on the same WSN node. In network level virtualization, subsets of nodes can form Virtual Sensor Networks (VSNs), dedicated to specific applications. This paper assumes a WSN that supports node level virtualization and studies the problem of assigning application sensing tasks to the physical and virtual sensors of the infrastructure.

Node level virtualization requires a virtualization software layer in the node, and this does lead to additional energy consumption. Energy consumption is an important factor to account for in WSNs and it must be minimized. In this paper, we consider a set of applications, with each application having multiple and different sensing tasks. The overall goal is to assign these tasks at the lowest cost in energy consumption. We consider the static version of the problem, i.e. given a set of application sensing tasks, we assign them to the physical and virtual sensors independently of their arrivals and departures. We formulate the problem as an Integer Linear Programming (ILP) problem and solve it in different scenarios, using IBM ILOG CPLEX. We compare the results to the case of a traditional WSN, i.e. with no support for node level virtualization. We show that significant energy can be saved with an adequate assignment of tasks in a WSN that supports node level virtualization.

The rest of the paper is organized as follows. Section II discusses the requirements, and then, Section III presents the related work. Section IV covers the system model and problem formulation. We provide the simulation setup and evaluation results in Section V, and conclude the paper in Section VI.

## II. REQUIREMENTS

Let us consider a set of sensor nodes, with different sensing capabilities, are deployed in a geographical area. Multiple applications, with different sensing tasks, can be considered there, in conjunction with the sensing capabilities. Assuming node level virtualization (as we do in this paper), there are two possibilities for assigning any given task. One approach is to assign it to a physical sensor. In this case, the sensor node will only execute that task. The second possibility is to assign it to a virtual sensor running on a physical sensor. The sensor node will then execute the task in parallel with other tasks that run on other virtual sensors instantiated on it, but with an energy consumption overhead due to the virtualization. In this context, two main questions arise. *First, with multiple sensor nodes capable of executing a specific sensing task, to which one should the task be assigned? Second, considering a WSN with node level virtualization, should the sensing task be assigned to a physical sensor or to a virtual sensor?*

Several requirements influence the solution. The first requirement is that the solution should enable concurrent applications over a node level virtualization enabled WSN, with each application having different sensing tasks. This allows to prevent the deployment of redundant WSNs. The second requirement is efficiency in energy consumption. Sensor nodes operate using batteries with limited power. It is therefore critical to manage their available energy properly to extend the lifetime of the WSN. The third requirement is to account for the integrality of expected applications sensing tasks, when making assignment decisions. This allows to derive a better solution than one derived with an incomplete knowledge of expected applications sensing tasks. Considering all of these requirements will ensure the efficient utilization, optimized energy consumption and availability of resources.

### III. RELATED WORK

The problem of task assignment in WSNs has received significant attention over the past years. Research efforts can be classified into two categories. The first category is that of static sensing task assignment in WSNs. Given a set of applications with sensing tasks, the aim is to optimize their assignment to sensor nodes in the WSN. The second category is dynamic sensing task assignment in WSNs. There, the aim remains to optimize the applications' sensing task assignment to sensor nodes in a WSN; however, the dynamic arrivals and departures of sensing tasks are considered.

#### A. Static Task Assignment

Many works study the static task assignment problem, specifically in WSNs that do not support virtualization. In [3, 4], the problem is formulated, with the objective of maximizing the number of applications to run. The same problem is investigated in [5] and [6], with the objectives of maximizing profit and maximizing satisfaction rate of tasks, respectively. These works are of little relevance to our work since they do not consider node level virtualization-enabled WSNs. Moreover, they do not discuss energy consumption.

In contrast to the above studies, the work in [7] accounts for energy consumption. However, it still targets traditional WSNs with no virtualization. The work in [8] studies the static task assignment problem in a WSN with virtualization. Yet, it considers a WSN with network level virtualization. Instead, in our work we study the problem of static task assignment in WSNs with node-level virtualization.

#### B. Dynamic Task Assignment

The dynamic sensing task assignment problem has been covered by multiple studies in traditional WSNs. In [4] the authors investigated the problem considering the tasks arrivals and departures. The authors in [9] propose an algorithm to allow sensing tasks to be assigned dynamically to sensor nodes. However, these works still do not consider concurrent applications and do not consider energy consumption. In [10], the authors introduce an algorithm to solve the dynamic task assignment problem, with the objective of enhancing energy consumption. The study in [11] proposes an analytical model and a distributed heuristic to address the problem of task assignment with energy harvesting capabilities. However, the work still does not consider concurrent applications. In [12-14], dynamic task assignment in virtualized WSNs is studied, with the objective of optimizing energy consumption. However, those studies remain different from ours as they consider network level virtualization.

Overall, existing works have covered the problem of sensing tasks assignment in traditional WSNs and in WSNs with network level virtualization. However, none has considered the problem of sensing tasks assignments in a WSN with node level virtualization. In our work, we study this problem, in its static version, while aiming at minimizing the energy consumption.

### IV. SENSING TASK ASSIGNMENT PROBLEM

We now present the problem of sensing tasks assignment in WSNs with node level virtualization. We introduce our system model in Sec. IV.A and problem formulation in Sec. IV.B.

#### A. System Model

We define $I$ as a set of sensor nodes in a WSN, with node-level virtualization capabilities. We use $i \in I$ to represent one sensor node in $I$. We define $J$ as a set of independent application sensing tasks to execute. We use $j \in J$ to refer to one sensing task in $J$. As we focus on static assignment of sensing tasks, we disregard their arrival and departure times. Each sensing task $j \in J$ requires a sensing capability at a specific location. A sensor $i \in I$ is placed in one location and has different sensing capabilities allowing it to execute specific tasks, within its sensing range. We define a matrix $L$, such that $L_{ij}=1$ if $j \in J$ can be executed by sensor node $i$. Otherwise, $L_{ij} = 0$.

#### B. Problem Formulation

We formulate the static sensing tasks assignment problem as an ILP problem. For a task to run on a sensor node, it can either be executed on it directly over the Physical Sensor (PS) or on a Virtual Sensor (VS) running on the sensor node. We define the following decision variables accordingly:

$$x_{ij} = \begin{cases} 1, & \text{if task } j \text{ is assigned to PS on } i \\ 0, & \text{otherwise} \end{cases}$$

$$y_i = \begin{cases} 1, & \text{if sensor node } i \text{ is virtualized} \\ 0, & \text{otherwise} \end{cases}$$

$$z_{ij} = \begin{cases} 1, & \text{if task } j \text{ is assigned to a VS on } i \\ 0, & \text{otherwise} \end{cases}$$

**Energy Cost:** The total energy cost in our problem can be divided into two cost components as follows.

1) PS energy cost ($C^{PS}$): The PS energy cost represents the energy that is consumed by running tasks directly on PSs over sensor nodes. It is obtained as follows:

$$C^{PS} = \sum_{i \in I} \sum_{j \in J} x_{ij} L_{ij} E^{PS} \quad (1)$$

where, $E^{PS}$ is the energy consumption of running one sensor node.

2) *VS energy cost* ($C^{VS}$): The VS energy cost represents the energy cost that is consumed by executing sensing tasks on VSs over sensor nodes. It covers the energy consumption for running the physical sensor nodes in addition to a virtualization overhead for each VS. It is obtained as follows:

$$C^{VS} = \sum_{i \in I} y_i E^{PS} + \sum_{i \in I} \sum_{j \in J} z_{ij} L_{ij} E^{VS} \quad (2)$$

where, $E^{VS}$ is the energy consumption overhead incurred by creating a VS over a physical sensor node.

In our problem, we aim at enabling decisions at the lowest energy cost. We thus define our objective as follows:

$$\min(C^{PS} + C^{VS}) \quad (3)$$

**Constraints:** We consider the following constraints.

Each task should be assigned to either a PS or a VS, as indicated in equation (4):

$$\sum_{i \in I} L_{ij} x_{ij} + \sum_{i \in I} L_{ij} z_{ij} = 1 \; ; \; \forall j \in J \quad (4)$$

A sensor node can operate in a physical mode or in a virtualized mode. If a sensor node operates in virtualized mode, no task should be assigned to it on a PS. Also, if a sensor node is operating in physical mode, it can support maximum one task on its PS. We enforce these conditions with equation (5):

$$\sum_{j \in J} x_{ij} L_{ij} \leq (1 - y_i) \; ; \; \forall i \in I \quad (5)$$

A sensor node can support a maximum number of VSs, referred to as $M$. We enforce this condition with equation (6):

$$\sum_{j \in J} z_{ij} L_{ij} \leq M y_i \; ; \; \forall i \in I \quad (6)$$

Finally, we ensure that the tasks assignments do not exceed the available energy of the sensor $E_i$ through equation (7):

$$C^{PS} + C^{VS} \leq E_i \; ; \; \forall i \in I \quad (7)$$

## V. EVALUATION

This section presents our performance evaluation. We describe the evaluation scenarios and the obtained results.

### A. Evaluation Scenarios

We conduct our evaluations in three scenarios with a different number of sensor nodes and sensing tasks, as indicated in Table I.

Table I. EVALUATION SCENARIOS

|  | S1 | S2 | S3 |
|---|---|---|---|
| Area (meter) | 100x100 | 150x150 | 200x200 |
| # of Sensors | 10 | 15 | 20 |
| # of Tasks | 8 | 12 | 16 |

The sensor nodes and the sensing tasks are randomly distributed in the geographical areas. We consider that sensors have a sensing range of 30 m [3]. For each sensor, we select its available energy in the range of [1.9 J, 3.4 J] [7]. We consider that the energy consumption for one sensor node, operating a physical sensor $E^{PS}$, is 0.017 mJ [7] which is the energy consumption for the sensing and transmission of information. Further, we consider the energy consumption overhead resulting from the creation of a virtual sensor on a sensor node to be equal to 10% of $E^{PS}$. A sensor node can support up to four virtual sensors; a maximum determined based on experiments conducted in our lab with Advanticsys sensors [15].

### B. Results

The following results are obtained when solving the problem.

*1) Tasks Assignment Decisions:*

We present the outcomes of solving our problem in scenarios 1, 2 and 3, through geographical plots in Fig. 1. We observe there that assignments on PSs and VSs coexist. However, the assignment decisions depend on the location of the tasks in the system. In particular, in regions with higher density of tasks, the model tends to assign the tasks to VSs, as this implies the lowest energy consumption cost.

*2) Number of Sensor Nodes:*

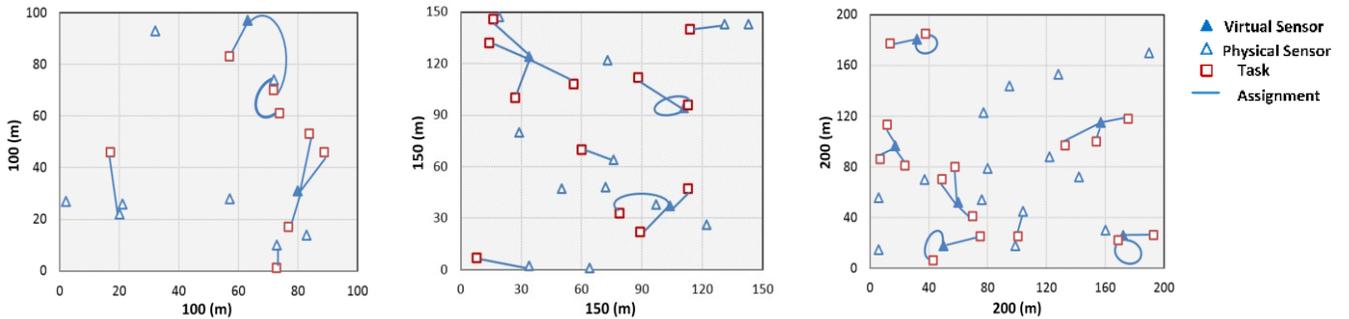

Fig. 1. Tasks assignment in Scenario 1 (left), scenario 2 (center) and scenario 3 (right).

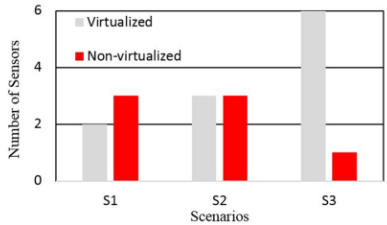

Fig. 2. Number of virtualized and non-virtualized sensor nodes in each scenario

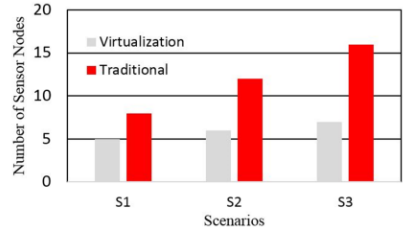

Fig. 3. Number of used sensor nodes in our problem vs traditional WSNs

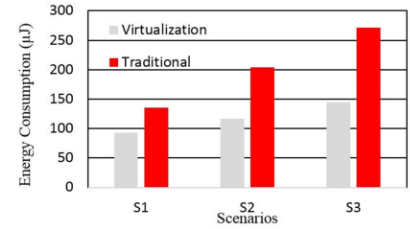

Fig. 4. Overall energy consumption in our problem vs traditional WSNs

We show in Fig. 2 the number of sensor nodes needed in each scenario. We distinguish between virtualized and non-virtualized sensor nodes. The figure confirms the previous observations: solving the problem leads to a combination of virtualized and non-virtualized sensor nodes.

Fig. 3 shows the number of sensor nodes resulting from our assignment in a virtualized WSN, over the three different scenarios. We compare this result to the number of sensor nodes needed in traditional WSNs, i.e. where each sensing task is assigned to one physical sensor. As can be observed, a traditional WSN can require up to two times more sensor nodes than our assignment in a virtualized WSN. This shows that with virtualization, solving the task assignment problem allows to optimize the number of operating sensor nodes in the WSN. This is especially important in regions with a high density of tasks, offering more flexibility in assignment decisions.

*3) Energy Consumption:*

We further evaluate the energy consumption based on our assignment in virtualized WSNs, considering the same three scenarios, in Fig. 4. This figure compares the overall energy consumption with our assignment in virtualized WSNs to that of traditional WSNs, i.e. where each sensing task is assigned to one physical sensor. As indicated, even though virtualization incurs the extra energy cost due to the overhead, we can still observe that an efficient assignment of tasks in virtualized WSNs can lead to up to 45% less energy consumption than with traditional WSN tasks assignments.

## VI. CONCLUSION

We studied the problem of sensing tasks assignment in WSNs, considering node level virtualization. We modeled the problem as an ILP problem, where we aimed at assigning sensing tasks to physical and virtual sensors, at minimum energy consumption cost. We evaluated our problem in different scenarios. Our results show that important savings in energy consumption can be obtained, compared to the traditional WSN. In the future, we plan to extend our work by considering dynamic arrivals and departures of sensing tasks, as well as different requirements of tasks in terms of sampling rates and priorities.